\documentclass[article,preprint,showpacs,superscriptaddress]{revtex4}
\DeclareRobustCommand{\baselinestretch{2}}

\usepackage{amssymb}
\usepackage{epsfig}
\usepackage{natbib}
\usepackage{amssymb}

\begin{document}

\title{Asymmetric optical radiation pressure effects on liquid interfaces under intense illumination}

\author{Alexis Casner}
\email{alexis.casner@cea.fr} \affiliation{Centre de Physique Mol\'{e}culaire Optique et Hertzienne,UMR
CNRS/Universit\'{e} 5798, Universit\'{e} Bordeaux I, 351 Cours de la Lib\'{e}ration, F-33405 Talence cedex, France}
\affiliation{Present address: D\'{e}partement Conception et R\'{e}alisation des Exp\'{e}rimentations, CEA-DAM Ile de
France, BP 12, F-91680 Bruy\`{e}res-le-Ch\^{a}tel, FRANCE}

\author{Jean-Pierre Delville}
\email{jp.delville@cpmoh.u-bordeaux.fr}

\affiliation{Centre de Physique Mol\'{e}culaire Optique et Hertzienne,UMR CNRS/Universit\'{e} 5798, Universit\'{e}
Bordeaux I, 351 Cours de la Lib\'{e}ration, F-33405 Talence cedex, France}

\author{Iver Brevik}
\email{iver.h.brevik@mtf.ntnu.no}
\affiliation{Department of Energy and Process Engineering, Norwegian University of Science and
Technology, N-7491 Trondheim, Norway.}

\date{\today}

\begin{abstract}
Deformations of horizontal liquid interfaces by optical radiation pressure are generally expected to display similar
behaviors whatever the direction of propagation of the exciting laser beam is. In the present experiment we find
this expectation to be borne out, as long as the cw laser illumination is moderate in strength. However, as a
striking contrast in the case of high field strengths, we find that either a large stable tether can be formed, or
else that a break-up of the interface can occur, depending on whether the laser beam is upward or downward directed.
Physically, the reason for this asymmetry can be traced to whether total reflection can occur or not. We also
present two simple theoretical models, one based on geometrical optics, the other on wave optics, that are able to
illustrate the essence of the effect. In the case leading to interface disruption our experimental results are
compared with those obtained by Zhang and Chang for water droplets under intense laser pulses [Opt. Lett.
\textbf{13}, 916 (1988)]. A key point in our experimental investigations is to work with a near-critical
liquid/liquid interface. The surface tension becomes therefore significantly reduced, which thus enhances the
magnitude of the stationary deformations induced.
\end{abstract}

\maketitle 

\section{INTRODUCTION}

 Since the seminal work of Ashkin and Dziedzic \cite{Ash2} on the deformation of transparent liquid free-surfaces
induced by laser waves, the optical radiation pressure has been recognized as very appealing to locally manipulate
liquid interfaces. The first developments exploited essentially the interface bending for optical applications in
adaptative lensing \cite{kats}, formation of surface relief holographic gratings \cite{komissa87}, or control of
morphology-dependent resonances \cite{changmdr}. Recently, radiation pressure effects have received renewed interest
in connection with nano/bio-technologies, as a non-intrusive tool to probe microscopic surface properties of soft
materials including cell stretching \cite{kasprl,kasbiophys} or membrane and interface viscoelasticity
\cite{wangol,wang01,sakai01,sakai02}. However, the bending of a fluid interface by the optical radiation pressure is
generally weak, and all these experiments were thus essentially limited to the investigation of the linear regime in
deformation.

The "linear regime" concept here ought to be defined more precisely, as there are actually two kinds of
nonlinearities to be recognized. First, hydrodynamic nonlinearity occurs if the slope of the liquid surface bulge is
large. In this context it is instructive to consider the Zhang-Chang droplet experiment \cite{zhang}. These authors
presented impressive pictures of the surface shape distortions in the front and in the rear region of a
micrometer-sized water drop. When illuminated with a strong laser pulse, droplet disruption was observed at the
rear. If the equilibrium radius of the droplet is $a$ and the surface elevation is $h$, linear hydrodynamic wave
theory is known to apply if the transverse gradient of $h$ is much less than one. In the case of a 100 mJ pulse, the
calculation of the Zhang-Chang experiment (Fig. 2 in Ref. 12) gave under typical conditions $h/a=0.02$ at the front
and $h/a=0.3$ at the rear. Accordingly, hydrodynamic linear theory should be expected to apply at the front but is
obviously not valid at the rear.

 The other kind of nonlinearity that may occur is of electromagnetic nature, caused by a large value of the field strength.
Generally, the ratio between the third-order and the first-order susceptibilities is written as
$\chi^{(3)}/\chi^{(1)}= E_0^{-2}$, where $E_0$ is for most materials of order of $10^{11}$ V/m (see Ref. 13).
Consider again the Zhang-Chang experiment. In case of the 100 mJ pulse the incident intensity at the centerline of
the beam was $ 0.29 \, {\rm GW/cm^2}$, which corresponds to an electric field of about $10^7$ V/m. As the droplet
acts as a lens, the field in the rear region is higher, perhaps by a factor of 2 or more. However, on the whole we
find it unlikely that the field strength was strong enough to affect the susceptibility significantly in the
Zhang-Chang experiment. We will consider similar conditions here. To be precise: we will assume {\it hydrodynamic
nonlinearity}, but {\it electrodynamic linearity}.

 The background for the present work is the
following. Whereas the linear regime is well understood and also delineated experimentally \cite{prlgiant}, the
nonlinear behavior at strong laser illumination is still at an early stage. Some success has so far been achieved in
the theoretical description of the weakly nonlinear regime of deformation \cite{lai,brevikdrop}. However such a
scheme, based as it is on linear wave theory, is clearly insufficient to explain the very large asymmetry observed.
This brings us to the primary aim of the present research, namely to show the existence of these large deformations
experimentally, to explore the physical reason for the asymmetry, and finally to present a simple theoretical model
whereby the essence of the phenomenon can be described. To the best of our knowledge, this effect is observed for
the first time at steady state under cw laser excitation, i.e. in a situation particularly favorable for
quantitative investigations. Even if the asymmetry as such is not novel, we think that the very large magnitude of
it, is.

 We  choose to work experimentally with liquid surfaces that are initially flat.
Considering a laser beam coming either from above or from below on the meniscus we are thus able to analyze the
surface deformations without entering into the complications arising from the lens effect in the case of curved
surface - cf. the striking demonstration of the last-mentioned effect in the Zhang-Chang experiment.

 We point out that the distortion of the liquid surface depends in general on the combined effect of gravity and surface
tension. The equation of motion for the height $h(r)$, where $r$ is the radius in cylindrical coordinates, is given
at steady state by:
\\
\\
\begin{equation}
  (\rho_{1}-\rho_{2}) g h(r)  - \sigma  \frac{1}{r} \frac{d}{d r} \Big (\frac{r h^{'}(r)}{\sqrt{1 + h^{'}(r)^{2}}}
   \Big ) = \Pi (r)
\label{equa1}
\end{equation}
\\
\\
where $\rho_1 , \rho_2$ are the densities of the two liquids, $\sigma$ the surface tension and $\Pi(r)$ the optical
radiation pressure. In the linear regime of deformation, $h^{'}(r) \ll 1$ and the radiation pressure value is taken
at normal incidence (see for instance Eq.~(1) in Ref. 14). Then for an exciting beam in the $TEM_{00}$ mode, the
solution for $h(r)$ can be calculated using a Fourier-Bessel transform and contains the capillary gravity wave
frequency $\Omega(k)=\sqrt{\frac{\rho_1 -\rho_2}{\rho_1+\rho_2}gk+\sigma \frac{k^{3}}{\rho_1+\rho_2}}$  as a central
parameter (cf. also p. 171 in Ref. 16). However by comparing the relative effect of gravity (buoyancy) to surface
tension, one defines an optical Bond number Bo as $Bo=(\omega_0/l_C)^2$, where $\omega_0$ is the beam waist and
$l_C=\sqrt{\sigma/(\rho_1-\rho_2)g}$ is the capillary length of the interface. When $ Bo \ll 1 $, and it will be the
case in the following experiments, then gravity turns out to be negligible for the surface distortion. The
centerline height is in this case inversely proportional to the surface tension $\sigma$, and is thus considerably
enhanced when $\sigma$ is a small quantity. A closer discussion on this point is given in Ref. 14.

\section{EXPERIMENTS}

\subsection{Set-up}

 Experiments were performed in a water-in-oil microemulsion (stable suspension of surfactant-coated water nano-droplets,
called micelles, dispersed in an oil-rich continuum). Its composition and some of its characteristics have already
been described previously \cite{prlgiant}. For a temperature $T >T_{C}$, where $T_{C} \simeq 308 K$ is a critical
temperature, the mixture separates in two micellar phases $\Phi_{1}$ and $\Phi_{2}$ of different concentrations.
Since the density (resp. index of refraction) of water is larger (resp. smaller) than that of oil, the micellar
phase $\Phi_{1}$ of larger concentration is located below the low micellar concentration phase $\Phi_{2}$ while its
refractive index $n_{1}$ is smaller than $n_{2}$ of $\Phi_{2}$. The main advantages of our medium are the extremely
weak surface tension of the liquid meniscus separating the two phases (typically $10^{6}$ times smaller at $T-T_{C}
= 3 \; K$ than that of the water/air free surface) and its very low residual absorption at the wavelength used. As a
consequence interface deformations can easily be monitored by continuous laser waves without disturbing thermal
couplings or non-linear bulk effects \cite{ol01lens}. Moreover the vicinity of a critical point ensures the
universality of the observed phenomena, because our mixture belongs to the universality class (d=3, n=1) of the
Ising model \cite{prlgiant}. It allows also us to evaluate the experimental parameters of our system according to
the following scaling laws for the surface tension $\sigma$ and the density contrast of the two phases : $\sigma=
\sigma_{0}\Big (\frac{T-T_{C}}{T_{C}} \Big) ^{2\nu}$, with $2\nu = 1.26$ and $\sigma_{0}=10^{-4} \; J.m^{-2}$;
$\Delta \rho = \rho_{1} - \rho_{2}= \Delta \rho_{0} \Big (\frac{T-T_{C}}{T_{C}} \Big) ^{\beta}$, with $ \beta =
0.325 $ and $\Delta \rho_{0} = 285 \; kg. m^{-3}$. For the refractive index contrast $\Delta n = n_{1} - n_{2}$, as
the two phases $\Phi_{1}$ and $\Phi_{2}$ are of very close composition, we assume the Clausius-Mossotti relation to
be valid \cite{prlgiant}: $\Delta n \simeq \Big ( \frac{\partial n}{\partial \rho} \Big )_{T} \Delta \rho$ with
$\Big ( \frac{\partial n}{\partial \rho} \Big )_{T}= -1.22 \; 10^{-4} \; m^{3}.kg^{-1}$. The Table 1 gives the
values of these parameters, together with the values of the absorption and thermal coefficients of our medium.

  The experimental set-up is shown in Fig.1. The mixture is enclosed in a thermoregulated spectroscopic cell
and the temperature is chosen above $T_{C}$ to reach the two-phase equilibrium state. The bending of the
liquid-liquid meniscus is driven by a linearly polarized $ TEM_{00}$ cw $Ar^{+}$ laser (wavelength in vacuum,
$\lambda_{0}= 514 $ nm) propagating either upward or downward along the vertical axis. The beam is focused on the
interface by an objective lens (Leitz 10X, N.A. 0.25). In the following $P$ is the beam power and the beam-waist
$\omega_{0}$, evaluated at $\frac{1}{e^{2}}$, is adjusted by changing the distance between a first lens L (focal
length f=1 m) and the focusing objective.

\subsection{Results}

 Typical interface deformations induced for identical experimental conditions ($T-T_{C} = 3 \; K$ and $\omega_{0} =
5.3 \mu m $) are presented on Fig. 2 for upward and downward directed beams. Since $n_{1} < n_{2}$, the radiation
pressure acts downwards toward the less refractive medium, regardless of the direction of propagation of the laser
\cite{brevik79,sakai02}. Fig. 3 shows the variation of the centerline height $h_0 = h(r=0)$ versus beam power $P$ in
both cases. As expected and already observed in experiments \cite{prlgiant}, $h_0$ is linear in $P$ at low beam
power. This regime corresponds respectively to $P \leq 225 \; \mbox{mW}$ and $P \leq 300 \; \mbox{mW}$ in the two
examples presented. Then when $P$ increases, $h_0$ gradually deviates from linearity. The deformation switches from
the classical bell shape to a stable tether shape in the upward case (see the last four pictures on Fig. 2a)). The
behavior is radically different in the downward excitation case. The deformation suddenly looses stability and $h_0$
diverges above a well-defined onset power $P_{S}$ (see for instance the two last pictures at $P=P_{S}= 400 \; mW $
on Fig. 2b)). This instability gives birth to a liquid jet that self-traps the beam and emits droplets at its end
(Fig. 2c).

\section{THEORETICAL DISCUSSION}

\subsection{The Importance of Total Reflection; scaling relation for the onset power $P_{S}$}

Actually, the reasons for this symmetry breaking can be understood from a very simple physical argument. As the beam
propagates from the large to the low refractive phase in the downward case, total reflection of light at the
interface can be reached. When this occurs, there is a dramatic concentration of light energy towards the tip of the
bulge which consequently becomes unstable. We have already demonstrated that, under a wide range of scaling
conditions, the measured onset power of the instability leading to the liquid jet is in complete accordance with the
onset power $P_{S}$ defined by the condition of total reflection \cite{jets}:
\\
\\
\begin{equation}
  P_{S} =  \frac{1.121 \; \pi}{0.715 \; \sqrt{2}} \frac{n_{1}}{n_{2}} \Big (1 + \frac{n_{1}}{n_{2}} \Big )
  \frac{\sigma c}{n_{2}-n_{1}} \; \omega_{0},
\end{equation}
\\
\\
where $\sigma$ as before is the meniscus surface tension and c is the light velocity in vacuum. The advantage of
working with critical fluids is another time obvious here, as $P_{S}$ scales as $\frac{\sigma}{n_{2}-n_{1}} \propto
(T - T_{C})^{0.93}$ and therefore vanishes close to $T \simeq T_{C}$.

For the opposite direction of propagation, the induced tethers (up to 60 $\mu m$ on Fig. 2a)) are surprising. A
coupling still exists between the laser propagation and the deformation because the bulge acts as a soft lens
\cite{ol01lens}. However, in this case the beam is focused inside the deformation and no amplification mechanism
therefore occurs for the intensity experienced by the tip of the bulge. These unusual nonlinear shapes deserve
further theoretical investigations. To our knowledge, they remain so far unexplained.

\subsection{A Simple Two-dimensional Geometrical Optics Model}

We find it worthwhile to point out that it is possible to get a physical picture of the essence of the asymmetry in
the downward/upward cases of the laser beam, without having to take into account the complex circular geometry of
the tether. Before embarking on calculations, one may ask to what extent an approximate picture in terms of
geometrical optics could be adequate. The important parameter distinguishing between wave optics and geometrical
optics is, for a sphere of radius $a$,
\\
\\
\[ \alpha = 2\pi a/\lambda_0. \]
\\
\\
The distinction between the two cases is not sharp, but in practice it is usually safe to work in terms of
geometrical optics if $\alpha \ge 80$,  as discussed, for instance, in Ref. 12. Assuming the highly curved region at
the tip of the distortion to correspond roughly to a sphere of radius $a=10\, \mu$m we obtain $\alpha \simeq$ 120
for $\lambda_0=514$ nm. The geometrical optics picture should according to this estimate be quite safe. Even if the
radius becomes halved, the resulting value $\alpha \simeq 60$ is most likely to be sufficient to justify the use of
geometrical optics in our case.

 Consider then the following simple model. Replace the tether with a symmetrical two-dimensional {\it wedge},
with an opening angle $2\beta$ facing upwards. Assume that an incident vertical ray in the downward direction falls
from the optical dense medium (2) towards one of the wedge surfaces, gets reflected towards the second surface, and
is thereafter reflected back in the upward vertical direction (see Fig.~4). This symmetric ray pattern becomes
accomplished if we choose $\beta= 45 ^0$, and assign the same value to the angle of incidence $\theta_2$. The
surface pressure $\Pi $ caused by one single reflection/refraction is given by the generic formula \cite{borzdov}:
\\
\\
\begin{equation}
\Pi (\theta_i)=\frac{n_{i}I}{c}\cos^2\theta_{i} \left\{ 1+R-\frac{\tan \theta_{i}}{\tan \theta_{t}} T \right\},
\end{equation}
\\
\\
where the index i refers to incidence and t to transmission, $I$ being the incident intensity. This formula can be
derived either by integration of the normal component of the volume force density ${\bf f}=-\frac{1}{2}E^2{\bf
\nabla} \varepsilon$ across the boundary region of the dielectric, or alternatively by considering the normal
component of Maxwell's stress tensor directly \cite{brevik79}. The coefficients of reflection and transmission, $R$
and $T$, satisfy the condition $R+T=1$. They can be found, for instance, on p. 496 in Stratton's book
\cite{stratton41}. The expression Eq.~(3) holds regardless of the state of polarization of the beam. In practice, we
usually insert the expressions for $R$ and $T$ corresponding to the TM or TE polarizations.

Consider first the downward incident ray, for which $\theta_{i}=\theta_{2}$ and $n_{i}=n_2$. Without losing the
essence of the problem we can make the simplifying assumption that this ray corresponds to the onset of total
reflection. Then from Snell's law, $n_2/n_1=\sqrt{2}$. (We are thus choosing refractive indices different from those
actually used in our experiment; this only to make the argument as simple as possible.) The two reflections from the
wedge surfaces contribute with equal weight to the pressure, since in this case $R=1,\, T=0$. We  calculate the
vertical force coming from $\Pi (\theta_{2})$ over a length $l$ of the surface ($l$ reckoned from the cusp),
multiply by 2 because of the two surfaces, and define a mean vertical pressure $\bar{\Pi}_z (\downarrow )$ by
dividing with the effective cross section $l\sqrt{2}$ for the considered part of the beam:
\\
\\
\begin{equation}
\bar{\Pi}_z (\downarrow)=\frac{n_2 I}{c}.
\end{equation}
\\
\\
For the reverse case of an upward directed incident ray, assumed to have the same intensity $I$, there is no total
reflection to be taken into account. The angle of incidence, now called $\theta_1$,  is $45^0$  as before, whereas
now $n_{i}=n_1$.  Assuming for definiteness the polarization to be in the plane of incidence (TM wave), we calculate
$R=0.005,\, T=0.995$, and find that $\Pi (\theta_{1})=0.359\, n_1 I/c$. The corresponding mean vertical pressure
becomes
\\
\\
\begin{equation}
\bar{\Pi}_z(\uparrow)=\frac{0.359\,n_1 I}{c}.
\end{equation}
\\
\\
The pressure ratio is
\\
\\
\begin{equation}
\frac{\bar{\Pi}_z(\downarrow)}{\bar{\Pi}_z(\uparrow)} \simeq 4,
\end{equation}
\\
\\
showing that the present crude model is able to predict a significant higher radiation pressure when the beam is
downward directed. It is therefore physically understandable that strong-field effects like the emission of droplets
occur in the downward case, i.e. when incidence occurs from the optically denser medium.

\subsection{Axially Symmetric Wave-Theoretical Model}

\bigskip

Quite justifiably, one may argue that the above model gives an over-simplified description of what happens
physically. The real situation is after all axially symmetric. To give a detailed wave-optical description of the
free surface displacement as a function of radius in cylindrical coordinates, would be rather complicated. Even the
description of the simple case of a monochromatic wave propagating inside a cylindrical medium of {\it infinite
length} is quite complicated (cf., for instance, Section 9.15 in Stratton's book \cite{stratton41}). However, it is
possible to get a grasp on the real physics by utilizing the known theory for a plane laser beam interacting with a
dielectric {\it sphere}, whose refractive index relative to the surroundings is greater than unity. Light becomes
concentrated at the rear region of the sphere as a consequence of the lens effect, resulting in an enhanced
electromagnetic energy density, and thus an enhanced surface force density, at the rear.

To illustrate this phenomenon mathematically, let us consider the simplified case of an isolated isotropic sphere of
radius $a$ and refractive index $n_2$ situated in an ambient medium of refractive index $n_1$,  illuminated by a
plane wave incident from above, along the $z$ axis. Taking the wave to be polarized in the $x$ direction, we can
represent it in complex notation as
\\
\\
\begin{equation}
{\bf E}^{(i)}=E_0 \,{\bf e}_x\,e^{ik_{1}z}, \quad {\bf B}^{(i)}=\frac{n_1 E_0}{c}{\bf e}_y\, e^{ik_{1}z}, \label{1}
\end{equation}
\\
\\
where $E_0$ is the amplitude and $k_1=n_1\omega /c$ the incident wave number. The time factor $\exp (-i\omega t)$
has been omitted. We shall need the electric field components on the interior surface of the sphere (superscript
$w$), at $r=a^-$. In standard notation the components can be written \cite{brevikdrop}
\\
\\
\begin{equation}
E_r^{(w)}(a)=\frac{E_0\cos \phi}{(k_2a )^2}\,\sum_{l=1}^\infty i^{l+1}(2l+1)\,a_l^{(w)}\,\psi_l(k_2a)P_l^1,
\label{2}
\end{equation}
\\
\begin{equation}
E_\theta ^{(w)}(a)=-\frac{E_0\cos \phi}{k_2a}\sum_{l=1}^\infty \frac{i^l(2l+1)}{l(l+1)}\left[
b_l^{(w)}\psi_l(k_2a)\frac{P_l^1}{\sin \theta}-ia_l^{(w)}\psi_l' (k_2a)\frac{dP_l^1}{d\theta}\right], \label{3}
\end{equation}
\\
\begin{equation}
E_\phi ^{(w)}(a)=\frac{E_0\sin \phi}{k_2a}\sum_{l=1}^\infty \frac{i^l(2l+1)}{l(l+1)} \left[
b_l^{(w)}\psi_l(k_2a)\frac{dP_l^1}{d\theta}-ia_l^{(w)}\psi_l' (k_2a)\frac{P_l^1}{\sin\theta}\right], \label{4}
\end{equation}
\\
\\
where $\psi_l(x)=xj_l(x)$ is one of the Riccati-Bessel functions, $k_2=n_2\omega /c$, and $a_l^{(w)}, b_l^{(w)}$ are
coefficients to be determined from the boundary conditions at $r=a$. These expressions are complicated and will not
be reproduced here. The local surface force density, repulsive when $n_2>n_1$, becomes
\\
\\
\begin{equation}
\Pi(\theta, \phi)=\frac{\epsilon_0}{2}(n_2^2-n_1^2)\left({E_t^{(w)}}^2+\frac{n_2^2}{n_1^2}
{E_r^{(w)}}^2\right)_{a^-}, \label{5}
\end{equation}
\\
\\
with ${E_t^{(w)}}^2 = {E_\theta^{(w)}}^2+{E_\phi^{(w)}}^2$. By means of the expansions (\ref{2}) - (\ref{4}) we can
calculate $\Pi(\theta,\phi)$ explicitly. It is convenient to write the force in the form of a series:
\\
\\
\begin{equation}
\Pi(\theta, \phi)=\frac{\epsilon_0}{2}(n_2^2-n_1^2)E_0^2\sum_{l=0}^\infty \sum_{m=-l}^l F_{lm}P_l^m(\cos
\theta)e^{im\phi}, \label{6}
\end{equation}
\\
\\
where the constant coefficients $F_{lm}$ can be calculated after inversion of the series. (The $l=0$ term descibes a
uniform pressure balance within the sphere.)

As one should expect from the form of the series (\ref{6}), the surface force density becomes most pronounced for
small or moderate values of the polar angle $\theta$ (the backward direction corresponds to $\theta=0$). Actually,
one can make the local surface force effect visible in practice, if one goes one step further and calculates the
hydrodynamic displacement of the spherical surface. This calculation was performed in full in
Ref.~\cite{brevikdrop}, for the case of a water sphere situated in air, and was shown to lead to a considerable
displacement at the rear end. The experiment of Zhang and Chang \cite{zhang} clearly showed the reality of the
effect. Beautiful computer-generated illustrations of the concentration of radiation energy at the rear of the
sphere can be found in the paper of Barton et al. \cite{barton88}.

Thus, to summarize, we have in the previous subsection, and in the present one, analyzed two different simple models
intending to give a rough description of the physics of our experiment. The first, 2D geometrical-optics ray-tracing
model, attributed the increased surface force  in the tip region (in the case of downward illumination) to the
presence of total internal reflection (Fig. 4). The second, cylindrically symmetric wave-optical model, attributed
the increased surface force to the lens effect for a sphere, together with the diffraction at the rear end. Neither
of these models are complete. However, they share the following important property in common with the real
experiment when illumination is taking place from above: Light incident along the symmetry axis in an optically {\it
dense} medium is refracted or diffracted into an outer optically {\it thin} medium through a convex surface. This
leads in all cases to an increased surface force, and is the essential physics of the effect.

\subsection{Dependence on Polarization, and on Angle of Incidence}

Another remark is called for, as regards the effect of different states of polarizations of the incident beam.
Usually, when dealing with this kind of situations  one tacitly assumes that the incident beam itself, as well as
the surface distortion, are azimuthally symmetrical. This is evidently true, if the beam is either nonpolarized or
circularly polarized. However, if the beam is linearly polarized, the azimuthal symmetry becomes lost. To what
extent does then the lack of symmetry in the case of linear polarization influence the distortion of the surface?

To illustrate this point, we show in Fig.~5 the normalized radiation pressure $\frac{\Pi (\theta_{1})} {\Pi (0)}$ on
a flat surface versus the angle of incidence $\theta_1$, when the beam falls from an optically thin ($\Phi_{1}$ in
our experiment) against an optically thick medium ($\Phi_{2}$). The curves, calculated separately for the TM and TE
polarizations, follow from Eq.~(3) and the Fresnel relations. We see that the sensitivity with respect to shift in
polarizations is only moderate, the curves being in fact coincident. For sake of comparison, we show in Inset of
Fig~5 the case of a water/air interface, light being incident from air ($n_1=1$) to water ($n_2=1.33$). The
dependence on polarization is more visible: in the TM case, there is a small enhancement of the pressure near the
Brewster angle.

On the other hand, Fig. 6 shows how the optical radiation pressure varies with the angle of incidence, here called
now $\theta_2$, when the ray is reversed and falls from $\Phi_{2}$ to $\Phi_{1}$. Once again, the dependence upon
shift in polarizations is seen to be very small for our experimental conditions, compared to the case of a water/air
interface (see Inset). The most notable feature of the curves is the pronounced directional effect: near the angle
of total reflection, the pressure becomes enhanced by a factor of $(1+n_1/n_2)^2$, as compared with the case of
normal incidence. This effect is of course the same as that discussed in Secs. III.A and B, but now regarded from a
somewhat different angle, as being a direct consequence of Fresnel's equations. It should be noticed that this
enhancement of the optical radiation pressure under total reflection condition was exploited by Komissarova {\it et
al.} in their experiments \cite{komissa87}, and also in Ref. 22. Let us note that in our situation $n_{1}$ and
$n_{2}$ are very close, due to the vicinity of the critical point. Then, while the total reflection angle is shifted
towards larger values, the sensitivity to polarization is even smaller.

As a conclusion, it is justified to consider for our experiment that the force is cylindrically symmetric, even for
a linearly polarized incident beam. This moderate sensitivity with respect to shift in local polarizations (the
local states of polarizations are of course different if we move around in the azimuthal direction on the distorted
surface), can be compared with what is observed if the refractive medium is not a dielectric but instead a metal. In
the latter case, there is no polarization dependence at all.  We refer the reader to the accurate measurement of
Jones and Leslie \cite{jones78} and to the comprehensive treatment given in Jones' monograph \cite{jones88}. This
experiment has been discussed theoretically also in Ref. 16, Appendix. The dependence upon polarization is thus
entirely associated with the dielectric property of the medium and is absent if its permittivity goes to infinity.

\section{Comparison with the Zhang-Chang experiment}

    First of all, it ought to be emphasized that the use of a laser {\it pulse} instead of a continuous beam in this
experiment does not make any difference in principle. The surface forces act in the same way in the two cases. It
was observed that the deflections of the free surface took place well after the passage of the pulse, but the reason
for this is the long hydrodynamic response time. The basic time scale is in fact determined by the transit time of
sound, across the dimensions of the medium \cite{brevik79}.  Now, for a sphere of radius $a=50 \, \mu$m this is not
very long, less than 0.1 $\mu$s. However, the complete displacement of the surface involves a series of multiple
internal reflections of sound at the surface, responsible for a longer delay. In the experiment, the surface
velocity was found to be appreciable for 10 $\mu$s or more.

This point being specified, as universality is ensured by the criticality of our experimental medium, Eq.~(2) should
of course be applicable to a flat water/air interface also ($\sigma \simeq \; 70 \; mN/m$ at room temperature).
Then, despite the presence of curved surfaces, the formation of a liquid fountain at the rear face of the droplet in
the experiment of Zhang and Chang \cite{zhang} should be at least roughly explainable in terms of the same
instability mechanism. From Eq.~(2) we find $P_{S}$=29 kW when $\omega_0=100\, \mu$m.  Now, it turns out that both
the 100 mJ and the 200 mJ laser pulses used in the Zhang-Chang experiment overshoot this limit. They correspond
respectively to $P= 125$ kW and 250 kW (see Ref. 12).

When taken at face value, these numbers indicate that there is a considerable discrepancy between theory and
experiment in the Zhang-Chang case. However, some care ought to be taken here, in view of the fact that Zhang and
Chang made use of short laser {\it pulses}. Let us for definiteness consider the 100 mJ pulse, distributed over a
cross-sectional disk of radius 100 $\mu$m. It is physically most appropriate to discuss the time-dependent
centerline intensity $I(t)$, rather than the total power $P$. The pulse can be modeled as  $I(t)=(I_0\,
t/\tau)\exp(-t/\tau)$, where $\tau=0.40\,\mu$s \cite{brevikdrop}. When the pulse is plane, $I_0$ is constant. With
$A=\pi \times 10^{-4}\, {\rm cm}^2$, we obtain $I_0=0.80\, {\rm GW/ cm}^2$. The maximum intensity,
$I_{max}=I_0/e=0.29\,{\rm GW/cm}^2$, occurs at $t=\tau$. Taking the effective duration of the pulse to be $3\tau$
($I(3\tau)=0.12\; {\rm GW/cm}^2$), the time-averaged intensity becomes $\bar I=0.21\,{\rm GW/cm}^2$. When this is
compared with the intensity of a stationary plane wave with power $P=$ 29 kW distributed over the same area (this
corresponds to $I=0.09\,{\rm GW/cm}^2$), we see that the Zhang-Chang measurements  yield a factor of about 2 times
the value predicted by Eq.~(2). One might here argue, however, that it is physically more appropriate to compare the
Zhang-Chang mean power of 0.21 GW/$\rm{cm}^2$ not with the case of a plane wave, but instead with a Gaussian beam at
a point located off the symmetry axis by a distance corresponding to one half of the centerline intensity maximum. A
Gaussian beam can be modeled, in the waist plane, as $I(r,t)=I(r)T(t)$, where
\\
\\
\begin{equation}
 I(r)=\frac{2P}{\pi \omega_0^2}\, \exp{(-2 r^2 /\omega_0^2)}, ~~~~  T(t) = \frac{t}{\tau}\exp{(-t/\tau)}.
\end{equation}
\\
\\
(The half-maximum intensity is seen to correspond to $r=0.59 \,\omega_0$.) According to this the Gaussian centerline
value should be doubled, from 0.09 to about 0.20 GW/${\rm cm}^2$, and the agreement with the Zhang-Chang experiment
becomes quite good.

The above estimate illustrates to what level we can expect agreement between theory and experiment. There are at
least three reasons why we cannot expect large accuracy. First, there is the roughness of the calculation. Secondly,
there is the fact that the original undisturbed surface in the Zhang-Chang experiment was already curved. Total
internal reflections could thus occur in that case. In the present case, without any distortion caused by the beam
itself, there will not be any total internal reflections. Finally, it is to be recalled that Eq.~(2) is dealing with
the limiting case of onset of total reflection only.  It is conceivable that some more power is required before the
large displacement of the surface develops in full. This is confirmed by dynamical investigations of the temporal
development of the instability \cite{unpub}. The last-mentioned point becomes applicable here because Zhang and
Chang reported the instability to occur in the region between 100 and 200 mJ.

\section{Final remarks and conclusions}
\subsection{On the Influence from Electrostriction}

When considering the electromagnetic forces above, we neglected the effect from electrostriction. To see the
legitimacy of this neglect more closely, let us briefly consider the expression for the full electromagnetic force
density in a nonmagnetic fluid (except from the fluctuating Abraham term)
\\
\\
\begin{equation}
{\bf f}=-\frac{1}{2}E^2 {\bf \nabla}\varepsilon +\frac{1}{2}{\bf \nabla}\left(E^2\rho \frac{d \varepsilon}{d\rho}\right),
\end{equation}
\\
\\
where $\rho$ is the mass density (cf., for instance, Refs.~12,16,20). Here the second term represents
electrostriction. Calculating this term by means of the Clausius-Mossotti equation, we get
\\
\\
\begin{equation}
{\bf f}=-\frac{\varepsilon_0}{2}E^2{\bf \nabla}\kappa+\frac{\varepsilon_0}{6}{\bf \nabla}[ E^2(\kappa-1)(\kappa+2)],
\end{equation}
\\
\\
$\kappa=\varepsilon/\varepsilon_0$ being the relative permittivity. The electrostriction pressure is always
compressive. The reason why it usually is left out in practical calculations, is that it does not contribute to the
total force on a test body. When this kind of force compresses the body, there is quickly established a
counterbalancing elastic pressure in the interior, so that the net influence on the body vanishes. The relevant time
scale for the establishment of this counterbalancing pressure is the time that sound needs to traverse the
illuminated region. For instance, if we estimate the transverse scale to be 10 $\mu$m we see, when taking the
velocity of sound to be 1500 m/s, that the relevant time scale becomes about 7 ns. Figure 9 in Ref.~16 shows, as a
result of a detailed calculation, how the elastic compensation in water becomes established in a very few
nanoseconds. The thesis of Poon \cite{poon90} may also be consulted for a very detailed analysis of these effects.
Our conclusion is thus that the electrostriction effect can safely be left out in the present case, as we are
dealing with stationary deformations.

\subsection{Conclusion}

The main purpose of the present work has been to experimentally show that Laser-Induced-Surface-Deformations become
asymmetric at high field strength. This asymmetry is characterized for the first time by using near-critical
liquid-liquid interfaces to strongly enhance optical radiation pressure effects. The mechanisms at the origin of
this asymmetry are presented and illustrated with very simple arguments. In particular, we demonstrate that the
dependence of optical radiation pressure versus the angle of incidence (Eq.~(3)) should be taken into account. This
point, which is generally neglected in the case of classical liquid interfaces, could be a first step to explain the
surprising tether shapes observed. Some preliminary calculations seem to confirm it \cite{mythesis}, even if the
numerical scheme used need to be refined.

In the case leading to interface instability, we also compare our experimental results with those obtained by Zhang
and Chang for water droplets under laser pulses excitation. This comparison is theoretically justified because
near-criticality leads to an universal description of radiation pressure effects. The predicted power onset turn out
to be of the same order than that observed with pulsed excitation.

 Finally, since the laser light propagates under total reflection condition inside the induced filament, this one provides,
to the best of our knowledge, the first example of non-permanent self-written liquid waveguide.

\begin{acknowledgments}
We are grateful to Max Winckert and J\"{o}el Plantard for technical assistance. This work was partly supported by
the CNRS and the Conseil R\'{e}gional d'Aquitaine.
\end{acknowledgments}

\newpage

\section*{Table}

\begin{table}[t]
\begin{center}
\begin{tabular}{|c|c|c|c|c|c|c|c|}

\hline $n_{0}$ & $\Delta n$ & $\rho_{0}\; (kg.m^{-3})$ & $\Delta \rho \; (kg.m^{-3})$ & $\sigma \; (J.m^{-2})$
& $\alpha_{th} \; (cm^{-1})$ & $\Lambda_{th} \; (W.cm^{-1}.K^{-1})$ & $D_{th} \; (cm^{2}.s^{-1})$ \\
\hline \hline
 1.464 & -7.6 $10^{-3}$ & 872  & 63.3  & 3 $10^{-7}$ & 3 $10^{-4}$ & 1.28 $10^{-3}$ & 8.8 $10^{-4}$ \\
\hline
\end{tabular}
\caption{Experimental parameters of our medium. $n_{0}$ and $\rho_{0}$: mean refractive index and density for $T <
T_{C} $. $\Delta n$, $\Delta \rho$, $\sigma$ : refractive index contrast, density index contrast and surface tension
evaluated at $T-T_{C}=3K$. $\alpha_{th}$, $\Lambda_{th}$ and $D_{th}$: optical absorption, thermal conductivity and
thermal diffusivity. }
\end{center}
\end{table}

\newpage

\section*{List of Figure Captions}

Fig. 1
\\
 Experimental set-up. BS: beam splitter, L: lens, M1,M2,M3: dielectric mirrors, $\frac{\lambda}{2}$:
 $\frac{\lambda}{2}$ plate, C: thermoregulated spectroscopic cell.
\\
\noindent Fig. 2
\\
 Interface deformations induced at $(T-T_{C})= 3 K$ by a laser beam of waist $\omega_{0} = 5.3 \; \mu m $.(a) Laser
propagating upwards from $\Phi_{1}$ to $\Phi_{2}$ as indicated by the white arrow. $P$ increases from top to bottom
and is successively equal to 210, 270, 300, 410, 530, 590 and 830 mW. (b) Downward direction of propagation. $P$ =
190, 250, 280, 310, 340, 370, 400 and 400 mW. The two last pictures are snapshots showing the destabilization of the
interface at $P_{S}$ leading to the formation of a stationary jet similar to that illustrated in (c) for $(T-T_{C})=
6 K$, $\omega_{0} = 3.5 \; \mu m $ and $P$ = 700 mW. $P_{S} = 490 \; \mbox{mW}$ in this last case. The total height
of picture (c) is 1 mm.
\\
\\
\noindent Fig. 3

 Evolution of the centerline height of the deformation $h_0=h(r=0)$ versus $P$ corresponding to the pictures of
Fig. 2a) ($\blacktriangledown$) and Fig 2b) ($\triangle$). Broken line indicates the onset $P_{S}$ above which the
interface becomes unstable when the laser is propagating downward.
\\
\\
\noindent Fig.4

Sketch of symmetric ray track when the beam is incident from above. Angle of incidence $\theta_2=\beta=45^0$;
$n_2/n_1=\sqrt{2}$.
\\
\\
\noindent Fig.5
\\
Normalized radiation pressure on flat surface versus angle of incidence $\theta_1$ for the TM and TE polarizations,
when $T - T_{C} = 3 K$ as in experiment. The light right is incident from $\Phi_{1}$ to $\Phi_{2}$. Inset: same
curves drawn for an air/water interface, ray incident from  air ($n_1=1$) to water ($n_2=1.33$).
\\
\\
\noindent Fig. 6
\\
Same as Fig. 5, but with direction of ray reversed, from $\Phi_{2}$ to $\Phi_{1}$, or for the Inset from water to
air. Angle of incidence is now denoted by $\theta_2$.

\newpage

  \begin{figure}[h]\centerline{\scalebox{1}{\includegraphics{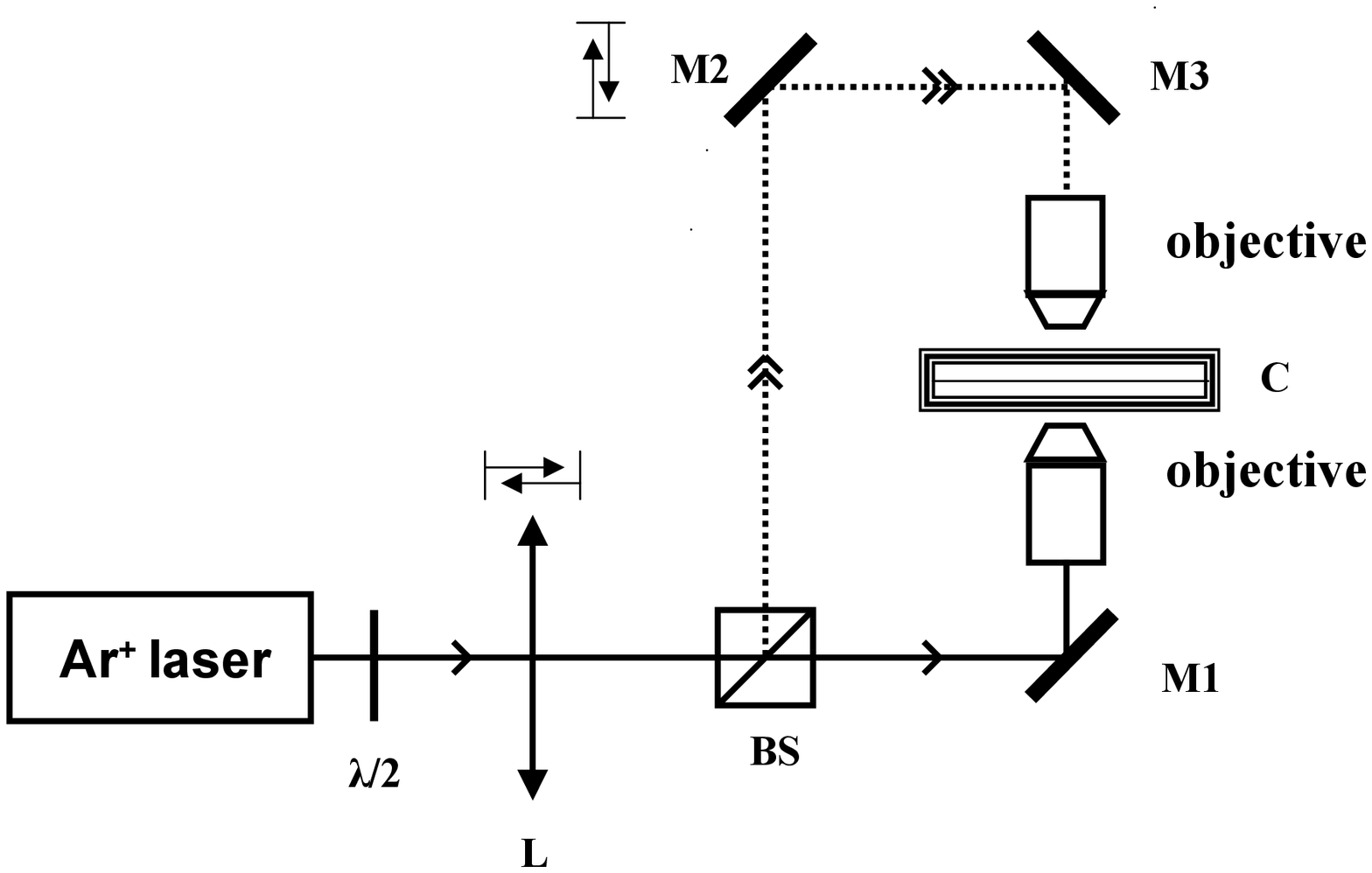}}}
  \end{figure}

\center{Fig. 1}

\newpage

 \begin{figure}[h]\centerline{\scalebox{1}{\includegraphics*{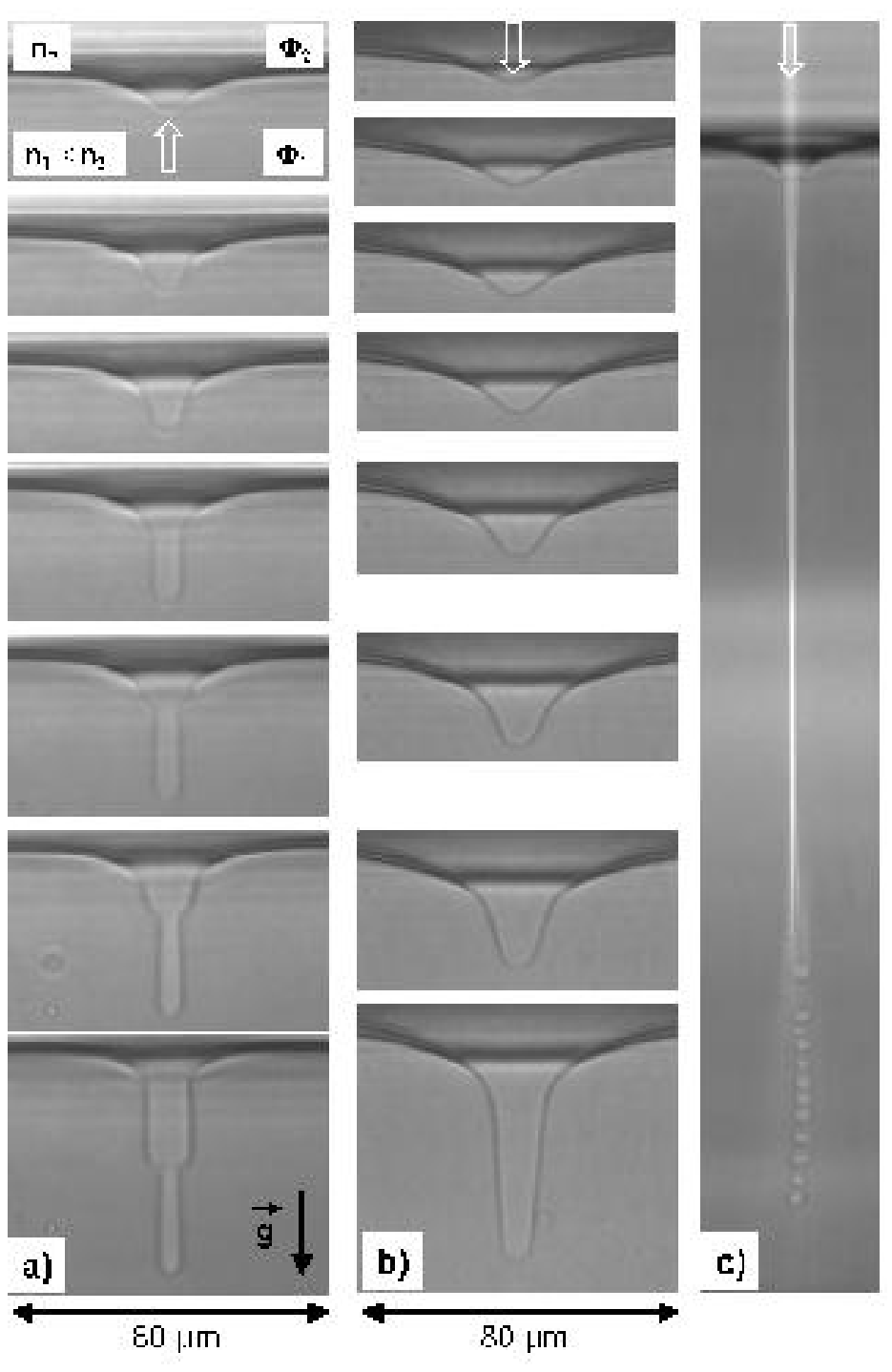}}}
  \end{figure}

\center{Fig. 2}

\newpage

  \begin{figure}[h]\centerline{\scalebox{1}{\includegraphics{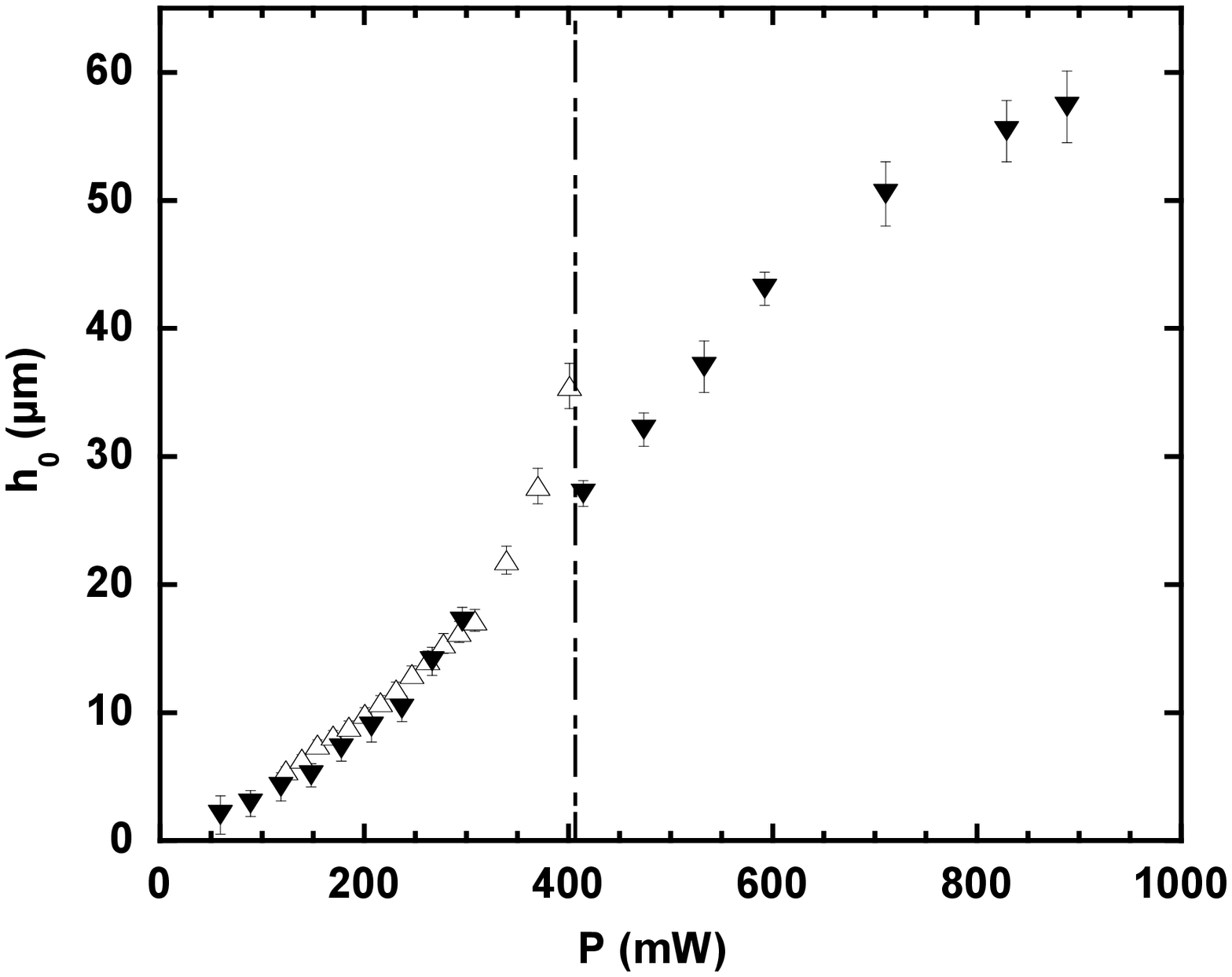}}}
  \end{figure}

\center{Fig. 3}

\newpage

  \begin{figure}[h]\centerline{\scalebox{1}{\includegraphics{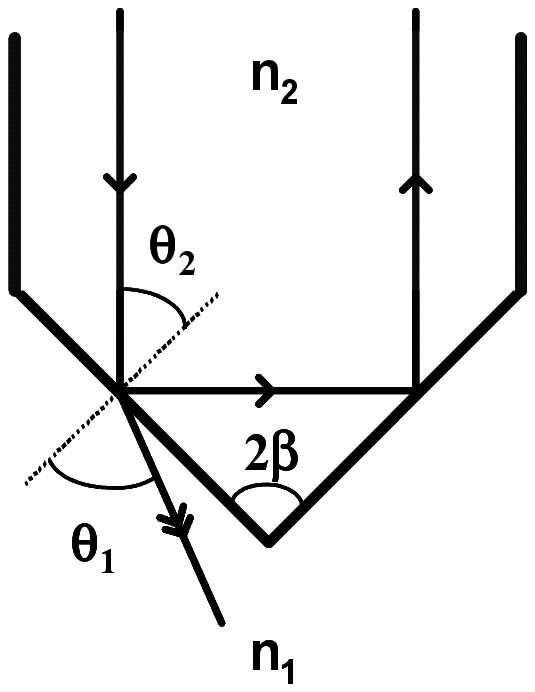}}}
  \end{figure}

\center{Fig. 4}

\newpage

  \begin{figure}[h]\centerline{\scalebox{1}{\includegraphics{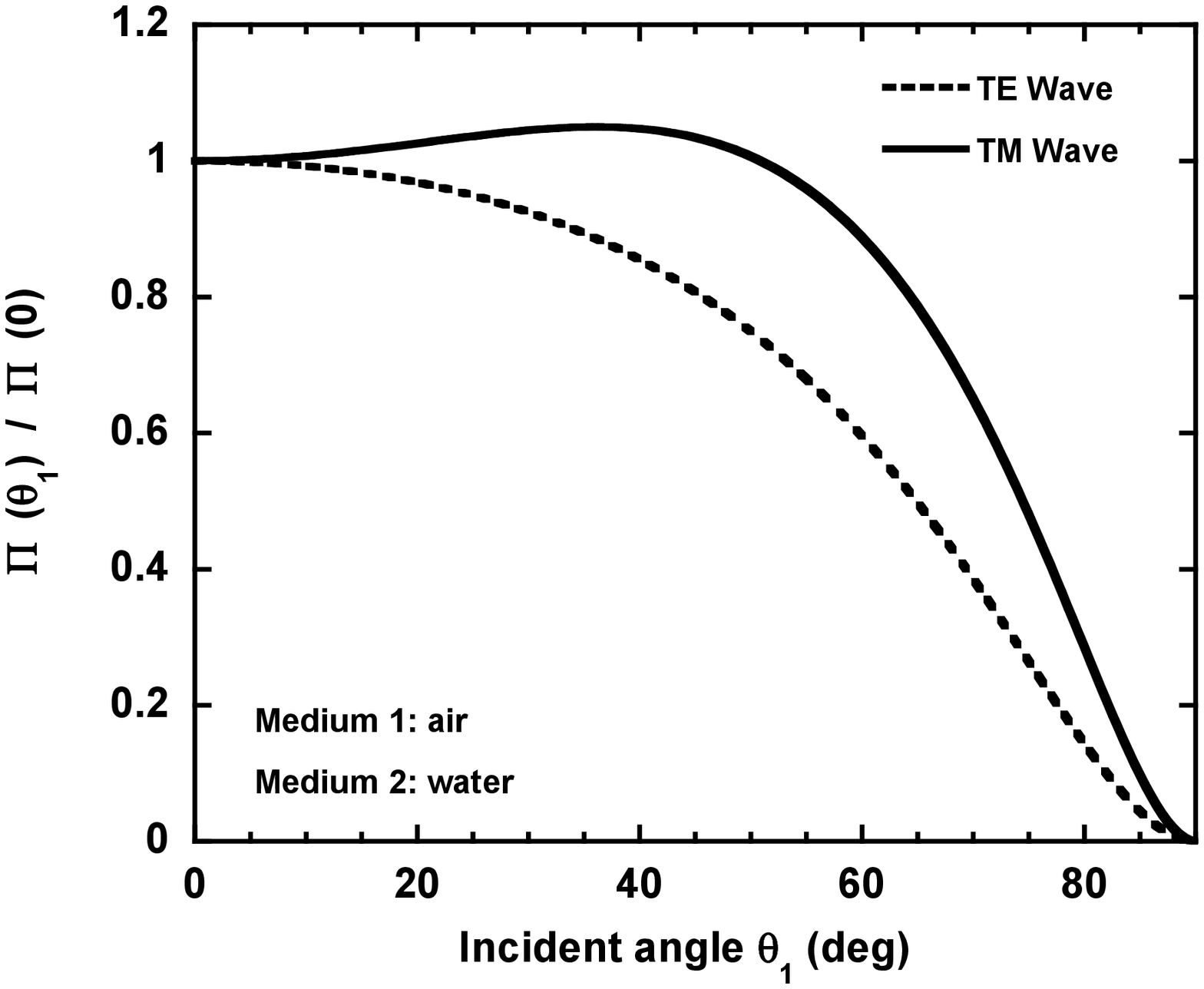}}}
  \end{figure}

\center{Fig. 5}

\newpage

  \begin{figure}[h]\centerline{\scalebox{1}{\includegraphics{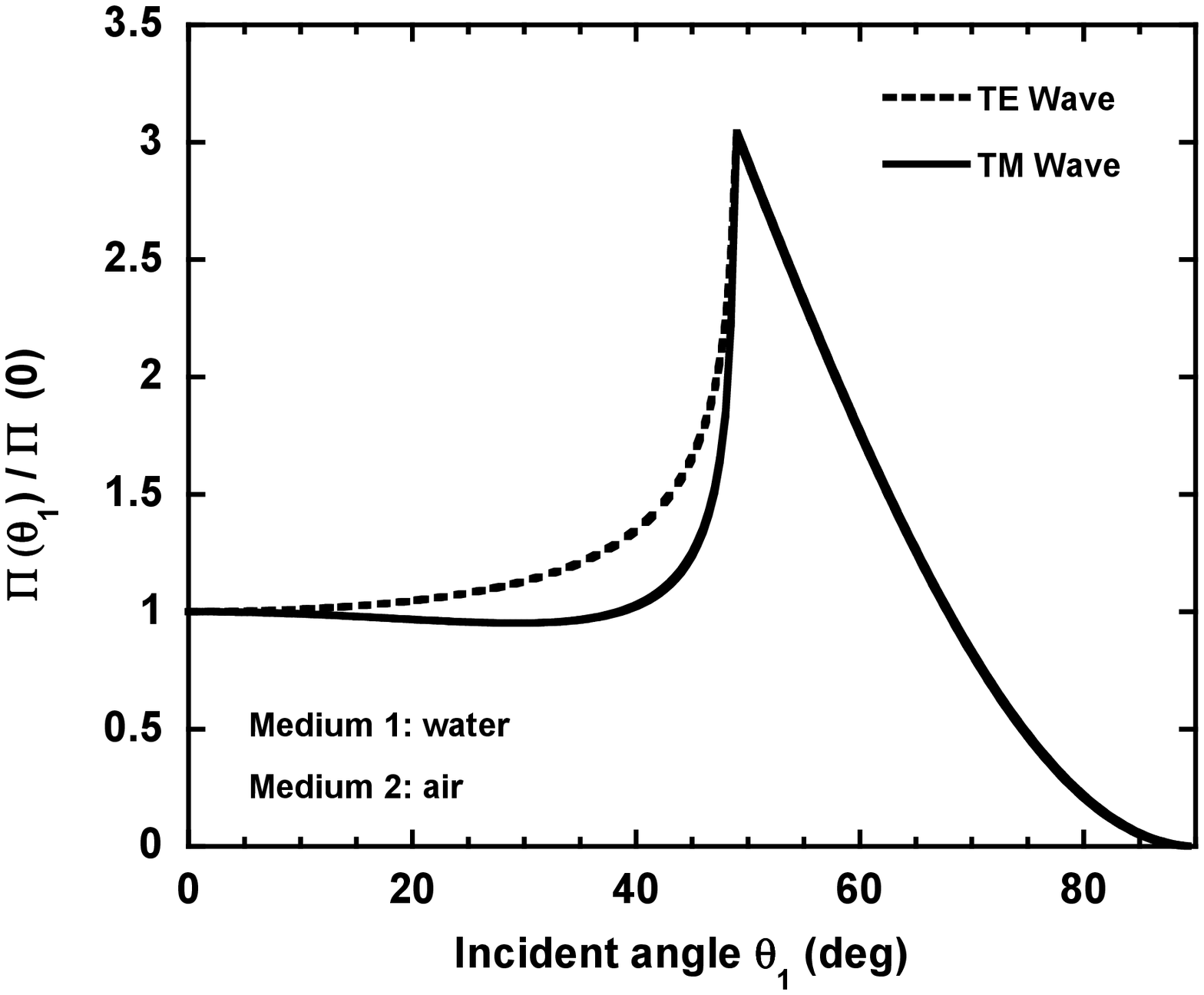}}}
  \end{figure}

\center{Fig. 6}

\end{document}